\documentclass[conference]{IEEEtran}
\usepackage{graphicx}
\usepackage{color}
\usepackage{multirow}
\usepackage{listings}
\usepackage{url}
\usepackage{balance}

\newif\ifdraft
\drafttrue
\ifdraft
\newcommand{\zhaonote}[1]{{\textcolor{cyan}    { ***Zhao:      #1 }}}
\newcommand{\note}[1]{ {\textcolor{red}    {\bf #1 }}}
\newcommand{\franknote}[1]{{\textcolor{green}    { ***Frank:      #1 }}}
\else
\newcommand{\zhaonote}[1]{}
\newcommand{\franknote}[1]{}
\newcommand{\note}[1]{}
\fi

\begin{document}
%

\title{Scientific Computing Meets Big Data Technology: An Astronomy Use Case}

%
%
%
%
%

\author{
\begin{tabular}{cccc}
{Zhao~Zhang$^{*,\circ}$} & {Kyle~Barbary$^{\circ,\dagger}$} & {Frank~Austin~Nothaft$^{*,\ddagger}$} & {Evan~Sparks$^*$} \\
{Oliver~Zahn$^\dagger$} & {Michael~J.~ Franklin$^{*,\circ}$} & {David~A.~Patterson$^{*,\ddagger}$} & {Saul~Perlmutter$^{\circ,\dagger}$}
\end{tabular}
\\
\begin{tabular}{c}
$^*$ AMPLab, University of California, Berkeley \\
$^\circ$ Berkeley Institute for Data Science, University of California, Berkeley\\
$^\dagger$ Berkeley Center for Cosmological Physics, University of California, Berkeley\\
$^\ddagger$ ASPIRE Lab, University of California, Berkeley \\
\end{tabular}
} 


\maketitle

\begin{abstract}
Scientific analyses commonly compose multiple single-process programs into a dataflow.
An end-to-end dataflow of single-process programs is known as a many-task application.
Typically, tools from the HPC software stack are used to parallelize these analyses.
In this work, we investigate an alternate approach that uses Apache Spark---a modern big data platform---to parallelize many-task applications.
We present Kira, a flexible and distributed astronomy image processing toolkit using Apache Spark. 
We then use the Kira toolkit to implement a Source Extractor application for astronomy images, called Kira SE.
With Kira SE as the use case, we study the programming flexibility, dataflow richness, scheduling capacity and performance
of Apache Spark running on the EC2 cloud.
By exploiting data locality, Kira SE achieves a 2.5$\times$ speedup over an equivalent C program when analyzing a 1TB
dataset using 512 cores on the Amazon EC2 cloud. 
Furthermore, we show that by leveraging software originally designed for big data infrastructure, Kira SE achieves competitive performance to the C implementation
running on the NERSC Edison supercomputer.
Our experience with Kira indicates that emerging Big Data platforms such as Apache Spark are a performant alternative for many-task scientific applications.  
\end{abstract}



\section{Introduction}
Dramatic dataset size increases have made data processing a major bottleneck for scientific research in many disciplines, such as astronomy, genomics, the social sciences, and neuroscience.
Researchers frequently start with a C or Fortran program that is optimized for processing a small amount of data on a single-node workstation and then use distributed processing frameworks to improve processing capacity. Examples include: the Montage astronomy image mosaic application~\cite{jacob09},  the sequence alignment tool BLAST~\cite{altschul90}, and high energy physics histogram analysis~\cite{ekanayake08}.
These applications are known as many-task applications because they comprise many small single-process tasks that are connected by dataflow patterns~\cite{raicu08}.

Scientists have used dedicated workflow systems (e.g., HTCondor~\cite{litzkow88}), parallel frameworks (e.g., the Message Passing Interface, MPI~\cite{gropp96}), and more recently the data processing system Hadoop~\cite{HADOOP}  to build these applications. Each approach has its own advantages such as provenance tracking, high scalability, and automated parallelism. However, these approaches also have shortcomings such as limited programming flexibility, lack of fault-tolerance, or a rigid programming model.

Apache Spark~\cite{zaharia12} was designed to support fast iterative data analyses on very large datasets.
Spark uses a directed acyclic graph (DAG) to describe parallel tasks, 
provides resilience against transient failures by tracking computational lineage
and optimizes for data locality when scheduling work.
These features make Apache Spark a compelling candidate distributed platform for many-task applications.

In this work, we investigate how to leverage Apache Spark,  as an example modern big data platform for many-task applications.
We study this question in the context of Kira, an astronomy image processing toolkit.
We use a real world application, an image source extractor (Kira SE), to examine the programming flexibility, dataflow richness,
and scheduling capacity of Spark. We evaluate Kira SE's performance by comparing against
an equivalent C implementation that is parallelized using HPC tools. Leveraging a platform like Apache Spark provides several advantages for many-task applications:
\begin{enumerate}
\item Spark can use existing astronomy libraries.
This allows astronomers to reuse existing libraries to build new analysis functionality.
\item Spark supports a broad range of dataflow patterns such as pipeline, broadcast, scatter, gather, reduce, allgather, 
and alltoall (shuffle). This broad dataflow pattern support can be used to optimize the data transfer between computation stages.
\item Spark's underlying file system support allows Kira to process data stored in a distributed file system such as HDFS~\cite{shvachko10}, 
as well as data stored in HPC-style shared file systems such as GlusterFS~\cite{davies13} or Lustre~\cite{donovan03}.
\item The Kira toolkit also inherits Spark's fault tolerance mechanism, which is a feature missing from MPI~\cite{gropp96}.
\end{enumerate}

Our experiments show that Spark is capable of managing $O(10^6)$ tasks and that Kira SE runs 2.5x faster than an equivalent C program when using a shared file system on the Amazon EC2 cloud with the 1TB Data Release 7 from the Sloan Digital Sky Survey~\cite{york00}. We also show that running Kira SE in the Amazon EC2 cloud can achieve performance that is competitive with running the equivalent C program running on the NERSC Edison supercomputer.

\newpage
Our experience with Kira indicates that Big Data platforms such as Apache Spark are a competitive alternative for many-task scientific applications.   
We believe this is important, because leveraging such platforms would enable scientists to benefit from the rapid pace of innovation and large range of systems and technologies that are being driven by wide-spread interest in Big Data analytics.
Kira is open source software released under an MIT license and is available from \linebreak \url{https://github.com/BIDS/Kira}.

\section{Background}
\label{sec:Background}

This section reviews the science behind sky surveys, introduces the source extraction operation, examines engineering requirements, and discusses the origin and usage of Spark.

\subsection{Sky Surveys}

Modern astronomical research is increasingly based around large-scale sky surveys.
Rather than selecting specific targets to study, such a survey will uniformly observe large
swaths of the sky. Example surveys include the Sloan Digital Sky Survey (SDSS)~\cite{york00},
the Dark Energy Survey (DES)~\cite{dark05}, and the Large Synoptic Survey Telescope~(LSST,
\cite{ivezic08}). Enabled by new telescopes and cameras with wide fields of view, these
surveys deliver huge datasets that can be used for many different scientific studies
simultaneously.

In addition to studying the astrophysical properties of many different individual galaxies,
the large scale of these surveys allows scientists to use the distribution of galaxies to
study the biggest contemporary mysteries in astrophysics: dark matter, dark energy, and 
the properties of gravity. These surveys normally include a time component: each patch of the sky is imaged many times,
with observations spread over hours, days, weeks or months. With this repeated imaging,
transient events can be detected via ``difference imaging''. Transients such as supernovae
can be detected in large numbers to better measure dark energy, and the large survey area
often results in the discovery of new, extremely rare, transient phenomena.

\subsection{Source Extraction}
\label{sec:Background-SE}
Source extraction is a key step in astronomical image processing pipelines.
SExtractor~\cite{bertin96} is a widely used C application for source extraction.
Although SExtractor is currently implemented as a monolithic C program for extracting astronomical
objects from images, the application's logic can be further divided into 
background estimation, background removal, object detection, and astrometric 
and photometric estimation.

Astronomers can improve extraction accuracy by running multiple iterations of source
extraction. Detected objects are removed after each iteration. While
the original C program contains the required functionality for building this iterative source extractor,
it does not expose the interfaces through the command line. To resolve this issue,
SEP~\cite{barbary2015} reorganizes the code base of SExtractor to expose core
functionality as a library. SEP provides both C and Python interfaces. Users can then
build the iterative source extractor with SEP's primitive. Kira SE is implemented by calling into
the SEP library.

\subsection{Engineering Requirements}
\label{sec:Background-EngReq}

In some experiments---such as experiments that look for supernovae explosions---it is important to process
the images as rapidly as possible. A rapid processing pipeline can enable astronomers to trigger follow-up
observations with more sensitive instrumentation before the peak of the supernovae occurs.
High throughput is also needed in large scale
sky survey pipelines that perform real time data analysis, e.g., LSST~\cite{ivezic08}.
The LSST uses a 2.4m-wide optical telescope that captures 3.2 billion pixels per
image. This telescope produces approximately 12.8~GB in 39~seconds
for a sustained rate of $\sim$330~MB per second. A typical night produces 13~TB of data. 
Over the planned 10-year project, the survey is expected produce 60~PB of raw data, which will be consolidated into a 15~PB catalog.
This timely processing requirement and the massive amount of data provides a challenging throughput requirement for the 
processing pipeline. 

\subsection{Spark}
Apache Spark is a dataflow-based execution system that provides a functional, collection
oriented API~\cite{zaharia12}. Spark's development was motivated by a need for a
system that could rapidly execute iterative workloads on very large datasets, as is common
in large scale machine learning~\cite{zaharia10}. Spark has
become widely adopted in industry, and academic research groups have used Spark
for the analysis of scientific datasets in areas such as neuroscience~\cite{freeman14} and genomics~\cite{nothaft15}.

Spark is centered around the Resilient Distributed Dataset~(RDD) abstraction~\cite{zaharia12}.
To a programmer, RDDs appear as an immutable collection of independent items that are 
distributed across the cluster. RDDs are immutable and are transformed using a
functional API. Operations on RDDs are evaluated lazily, enabling the system to schedule 
execution and data movement with better knowledge of the operations to be performed than 
systems that immediately execute each stage. Spark provides Scala, Java, and Python programming interfaces.
By default, Spark uses HDFS~\cite{shvachko10} for persistent storage, but
it can process data stored in Amazon S3 or a shared file system such as
GlusterFS~\cite{davies13} or Lustre~\cite{donovan03}. Spark provides fault tolerance via
lineage-based recomputation. If a partition of data is lost, Spark can recover the data
by re-executing the section of the DAG that computed the lost partition.


\section{Applying Spark to \\ Many-Task Applications}
\label{sec:Capability}

Scientific analysis pipelines are frequently assembled by building a dataflow out of many
single-process programs. Many-task applications arise in scientific research
domains including astronomy, biochemistry, bioinformatics, psychology, economics, climate science,
and neuroscience. The tasks in these applications are typically grouped into stages and the stages are connected by
producer-consumer data sharing relationships. 
A previous survey study~\cite{katz11} identified
seven common dataflow patterns among a group of many-task applications. The patterns
include pipeline, broadcast, scatter, gather,
reduce, allgather, and alltoall. Most many-task applications
can be viewed as stages of independent tasks that are linked by these dataflow patterns.

The map-reduce~\cite{dean04} model uses a similar pattern to schedule jobs. Traditional map-reduce
systems such as Google's MapReduce~\cite{dean04} and Apache Hadoop MapReduce~\cite{HADOOP} 
abstract producer-consumer relationships into a map stage and a reduce stage. These two stages are then
linked by a data shuffle. Although these systems have proved very powerful for processing very
large datasets, the map-reduce API has been criticized as inflexible~\cite{dewitt08}.
Additionally, since jobs are restricted to a single map and reduce phase, tools such as
FlumeJava~\cite{chambers10} are necessary for assembling pipelines of map-reduce jobs. Since
data is spilled to disk at the end of each map and reduce phase, traditional map-reduce platforms perform poorly
on iterative and pipelined workflows~\cite{zaharia12}.

To resolve these problems, second-generation map-reduce execution systems such as
DryadLINQ~\cite{yu08} and Spark~\cite{zaharia12} allow for applications to be decomposed into
DAGs. In these DAGs, nodes represent computation, and the nodes are
linked by dataflows. In Spark, this abstraction is provided by RDDs~\cite{zaharia12}.

In Table~\ref{tb:Patterns}, we demonstrate how seven common dataflow patterns can be mapped
to Apache Spark.

\begin{table}[h]
  \begin{center}
  \caption{Dataflow Pattern Primitives in Spark}
    \begin{small}
    \begin{tabular}{ | p{1.8cm} | p{5.5cm} |}
    \hline
    Pattern & Spark primitive \\
    \hline \hline
    Pipeline & RDD.map()  \\ 
    Broadcast & sparkContext.broadcast() \\   
    Scatter & sparkContext.parallelize() \\ 
    Gather & RDD.collect() \\ 
    Reduce & RDD.reduce() \\ 
    Allgather & RDD.collect().broadcast() \\ 
    Alltoall & RDD.reduceByKey() or \\
 & RDD.repartition() \\ 
    \hline
    \end{tabular}
    \end{small}   
  \label{tb:Patterns}     	
  \end{center}
\end{table}

Spark improves upon
Hadoop MapReduce by adding an in-memory processing model that natively supports iterative computation. 
As compared to other DAG based methods such as DryadLINQ, this enables
the efficient execution of chained pipeline stages. In a chained pipeline,
I/O and communication are only performed before the first stage of the chain, and after the
last stage of the chain. Spark uses communication barriers to synchronize the execution of each stage~\cite{zaharia12}.

\section{Kira Design and Implementation}
\label{sec:Design}

When designing the Kira astronomy image processing toolkit, 
we focused on improving computation and  I/O cost as well as on providing an flexible programming interface and supporting code reuse.

\subsection{Architecture Overview}
\label{sec:Design-Overview}

\begin{figure}[t]
	\begin{center}
		\includegraphics[width=85mm]{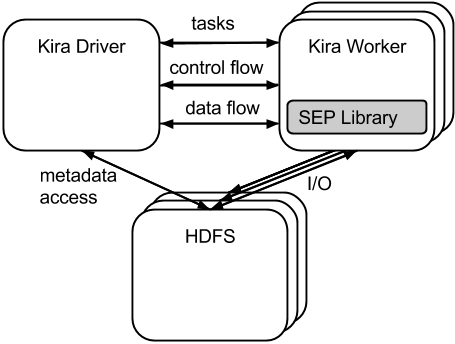}
		\caption{Overview of Kira's Architecture and Inter-Component Interactions}
		\label{fig:architecture}
  	\end{center}
\end{figure}

Kira's overall architecture is shown in Figure~\ref{fig:architecture}. 
Each outer box with rounded corners is a process. A process can be
a Kira Driver, a Kira Worker, or a HDFS daemon (NameNode or DataNode).
Kira runs on top of Spark, which supports a single driver and multiple workers.
The SEP library (shaded inner box) is deployed to all workers nodes. The input files are
stored in the underlying file system.

To run Kira, we dispatch the compiled program, the parameters, and library dependencies to the Kira Driver.
The Kira Driver is then responsible for managing control flow, dataflow, and task scheduling 
by coordinating the Kira workers. 
The Kira Driver accesses distributed/parallel file systems for metadata.
The actual I/O is distributed across the Kira worker nodes in
parallel. To run a task, workers use the Java Native Interface (JNI) to call the SEP library.

\subsection{Computation}
\label{sec:Design-Computation}
We considered three approaches when implementing the Source Extractor algorithm in Kira:

\begin{enumerate}
\item We can \textbf{reimplement} the Source Extractor algorithm from scratch.
\item We can connect existing programs as \textbf{monolithic} pieces without changing them.
\item We can reorganize the C-based SExtractor implementation to expose a programmable
\textbf{library} that we call.
\end{enumerate}

While \textbf{reimplementing} the functionality of the C SExtractor code using Spark's Scala API 
would allow us to execute SExtractor in parallel, it would lower the efficiency of the computation 
and would require a significant reimplementation effort.
The \textbf{monolithic} approach would not involve a modification to the original executable. While we could 
integrate with the original codebase at this level, this would lock us in to the hardcoded program logic 
of the original program. 
For example, astronomers can improve extraction accuracy by running multiple iterations of source
extraction with detected sources being removed from the input image after each iteration. 
The original SExtractor contains the required functionality for this iterative process, however, the hardcoded
logic only allows users to run the source extraction once.  
In order to not be locked in to the rigid control flow limitations of the \emph{monolithic} model, 
we instead opt for a \textbf{library}-based model.
This approach allows us to reuse the legacy code base without sacrificing control-flow flexibility.  


\subsection{I/O}
\label{sec:Design-I/O}
The Flexible Image Transport System (FITS)~\cite{wells81} format is a widely adopted file format for astronomy
images. Each FITS file contains ASCII metadata and binary image data.
The FITS format is commonly used by sky surveys, thus Kira must be able to process and export FITS files.
In Kira, one of our goals is to leverage the locality information provided by HDFS. When a file is loaded into 
HDFS, the file is split into blocks that are replicated across machines. When a system such as Spark loads the file, 
HDFS then provides information about which machines have each block of the file. This allows the
scheduler to optimize task placement for data locality.


Our solution uses the \texttt{SparkContext.binaryFiles()} API.
This API loads all files within a directory as a sequence of tuples. Each tuple contains the file object and a byte
stream containing the contents of the file. We then use the jFITS~\cite{jfits} library to convert these byte streams 
into the FITS objects that users can transform and compute upon.

\section{Programming Kira}
\label{sec:Programming}
The Kira API is described
in Table~\ref{tb:Primitives}. Background methods are used to estimate and remove the image background. The Extractor
API is used for extracting objects and estimating astrometric and photometric parameters. The Ellipse API offers helper
functions for converting between ellipse representations, and for generating masks that are based on an object's elliptical
shape. The \texttt{sum\_circle()}, \texttt{sum\_ellipse()}, and \texttt{kron\_radius()} methods in the extractor category and all
methods in the ellipse category perform batch processing, where the input coordinates are passed as a three dimensional array. By processing objects
in batches, we are able to amortize the cost of each Java Native Interface (JNI) call over many objects.

\begin{table*}[t]
\begin{center}
\caption{Kira Primitives and Explanation}
\label{tb:Primitives}
\begin{tabular}{ |l|l|l| }
\hline
Group & API & Explanation \\ \hline \hline
\multirow{3}{*}{Background} & makeback() & Builds background from an input image \\
 & backarray() & Returns the background as a 2D array \\
 & subbackarray() & Subtracts a given background from image \\ \hline
\multirow{4}{*}{Extractor} & extract() & Returns objects extracted from the input image \\
 & sum\_circle() & Sums data in circular apertures \\
 & sum\_ellipse() & Sums data in elliptical apertures \\ 
 & kron\_radius() & Calculate iron radius within an ellipse \\ \hline
\multirow{3}{*}{Ellipse} & ellipse\_coeffs() & Converts from ellipse axes and angle to coefficient representations \\
 & ellipse\_axes() & Converts from coefficient representations to ellipse axes and angles \\ 
 & mask\_ellipse() & Masks out certain pixels that fall in a given ellipse \\ \hline
\end{tabular}
\end{center}
\end{table*}

This API allows us to build a source extractor in Kira that is equivalent to the SEP
extractor~\cite{barbary2015}.
Listing~\ref{lst:SE} contains pseudocode describing how to implement a source extractor
using Kira's API. This code uses Spark's \texttt{binaryFiles()} method to load input
files from persistent storage. We then map over each file to convert the FITS data into
a matrix with associated metadata. In the final map stage, we estimates and remove the
background from the matrix. Once the background is removed, we then extract the objects
from the matrix.

\begin{lstlisting}[caption=Objects Extraction Logic, label=lst:SE, linewidth=0.5\textwidth, xleftmargin=2.5ex]
val input_rdd = sparkContext.binaryFiles(src)
val mtx_rdd = input_rdd.map(f => load(f))
val objects_rdd = mtx_rdd.map(m => {
  /* mask is a 2-d array with 
   * the same dimensions as m
   */
  val mask = null
  val bkg = new Background(m, mask)
  val matrix = bkg.subfrom(m)
  val ex = new Extractor
  val objects = ex.extract(matrix))
})
\end{lstlisting}

In Listing~\ref{lst:SE-Iter}, we demonstrate how the Kira API can be used to perform
iterative image refinement. Although the original SExtractor~\cite{bertin96} contains all necessary functionality, 
it is not feasible for users to implement this feature due to the hardcoded program logic . However, since Kira provides
library level bindings, it is easy to implement a multi-stage refinement pipeline.

\begin{lstlisting}[caption=Iterative Objects Extraction Logic, label=lst:SE-Iter, linewidth=0.5\textwidth, xleftmargin=2.5ex]
val input_rdd = sparkContext.binaryFiles(src)
val mtx_rdd = input_rdd.map(f=>load(f))
val objects_rdd = mtx_rdd.(m => {
  /*mask is a 2-d array with 
   *the same size of m
   */
  var mask = null
  var ex = new Extractor   
  for(i <- 0 until 5) {
    var bkg = new Background(m, mask)
    var matrix = bkg.subfrom(m) 
    var objects = ex.extract(matrix)
    mask = mask_ellipse(objects)  
  }
  objects
})
\end{lstlisting}

Listing~\ref{lst:SE-Iter} wraps the source extraction phase from Listing~\ref{lst:SE} in a loop. 
This allows us to update the mask used for extraction, which is used to further refine the extraction
 in subsequent iterations.

\section{Tuning Spark}
This section discusses how we configure Spark in terms of parallelism and scheduling to make
Kira SE more efficiently use EC2 computing resources. 

\subsection{Parallelism}

Spark allows for both thread and process parallelism. By default, Spark makes use of
thread-level parallelism by launching a single Java Virtual Machine~(JVM) per worker machine.
Users then specify the number of threads to launch per worker (typically, one thread per core).
However, in Kira SE, neither the jFITS nor the SEP libraries are thread safe. 
To work around this,
we configured Spark to support process level parallelism by launching a worker instance for each
core. This configuration may reduce scalability,
as it increases the number of workers the driver manages. However, our experiments with 512
workers in~\S\ref{sec:Performance} show that Kira's scalability is not severely impacted by
worker management overhead.

\subsection{Scheduling}

Spark's task-scheduling policy aims to maximize data locality by using
delay scheduling~\cite{zaharia10ds}. In this scheduling paradigm, if node $n$ has the data needed
to run job $j$, job $j$ will execute on node $n$ if job $j$ would wait less than a threshold time
$t$ to start. The policy is tunable through three parameters:

\begin{itemize}
\item{spark.locality.wait.process}
\item{spark.locality.wait.node}
\item{spark.locality.wait.rack}
\end{itemize}

These parameters allow users to specify how much time a task will wait before being sent to another
process, node, or rack. For Kira SE, we have found that data balancing can 
impact task distribution, leading to node starvation and a reduction in overall performance.
Loading a 65~GB (11,150 files) dataset from SDSS Data Release 2 to a 16-node HDFS deployment ideally should result in 699 files on each node.
In reality, the number of files on each node varies between 617 and 715.
Enforcing locality with longer {\em spark.locality.wait} time (3000 ms) leads to task distribution imbalance, which makes
Kira SE 4.5\% slower than running with {\em spark.locality.wait} set to 0~ms.
In practice, we set all {\em spark.locality.wait} parameters to zero, so that tasks do not wait for locality.
This setting effectively avoids starvation and improves the overall time-to-solution.

The root cause of the ineffectiveness of delay scheduling is the size of the input files for the Kira SE tasks. 
Each input file is $\sim$6~MB, compared to a typical block size of 64/128~MB~\cite{shvachko10}.
Delay scheduling's parameters let users specify how long a task should wait for locality before getting executed elsewhere.
This waiting time can be viewed as the expected job completion time difference between executing the task with data locality and without locality.
In the Kira SE case, the need for scheduling a task to a node without locality only occurs when there is a possible
starvation if we continue to enforce the locality. Since our input files are small, the cost of doing a remote fetch is low and thus we should only
wait for a short period time~\cite{ananthanarayanan11}. By sweeping the waiting time, we found that not waiting 0ms delivered the best performance.
While this result seems to contradict our result in~\S\ref{sec:Performance} that states that Apache Spark outperforms
the HPC solution due to data locality, it is not contradictory because even with no delay scheduling, 
98\% of the tasks scheduled have local data.  Thus the delay scheduling penalty is not needed in this case.   
We have traded a 2\% decrease in locality for a 4.5\% improvement in overall performance.

\section{An HPC Solution}
\label{sec:HPC}
\begin{figure}[t]
	\begin{center}
		\includegraphics[width=85mm]{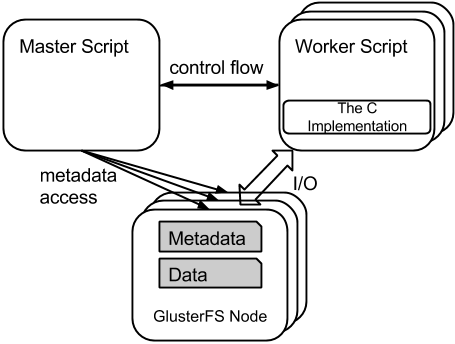}
		\caption{Overview of a Parallel Version of Source Extractor using HPC Tools}
		\label{fig:hpc-architecture}
  	\end{center}
\end{figure}

A typical way to parallelize a many-task application like source extractor is to use a scripting language
or MPI to launch multiple tasks concurrently. Figure~\ref{fig:hpc-architecture} shows the architecture
of such a solution.

All the input files are stored in the shared file system, e.g., GlusterFS or Luster. We use a master script
first to read all the input file names, then partition the file names into partitions. After that, the master
script informs the worker scripts on each node to process an independent partition of files in parallel 
in a batch manner.

For GlusterFS, all metadata is spread across all nodes. So the metadata query from the master script
needs to communicate with all nodes. The file I/O is done by the worker nodes. Note that unlike HDFS, 
POSIX I/O libraries cannot take advantage of data placement locality during file read/write. While Lustre 
file I/O is similar to GlusterFS, file system metadata is consolidated onto a small set of metadata servers.

\section{Performance}
\label{sec:Performance}

We evaluate Kira SE's performance against the HPC solution with the SEP library
(referred to as the C version in the following text). Because it uses SEP, Kira SE performs an identical amount of
computation and I/O as the C version.

To understand Spark's overhead, we compare the C implementation against Kira SE on a single
machine. Then we fix the problem size and scale Kira SE and the C implementation across a
varying number of machines on EC2 to understand the relative scalability of each approach.
The 1TB experiments demonstrate the difference in performance between Kira SE and the C version
for large dataset processing. Finally, we show some interesting results when running the C
version on the Edison supercomputer that is deployed at the National Energy Research Scientific Computing Center (NERSC).

In all experiments using Amazon's EC2 service, we use m2.4xlarge instances.
This instance type has eight cores (each running at 2.4GHz), 68~GB RAM, two hard disk drives (HDD), and Gigabit Ethernet. 
We chose a HDD based configuration as opposed to Amazon's newer solid state drive (SSD) backed instances to give a fair comparison
against the Edison supercomputer, which is backed by HDDs. 
Software configurations are described in the following experiments.

\subsection{Single Machine Performance}
\label{sec:Performance-scaleup}

The purpose of the scale-up experiments is to understand Spark's overhead by running
Kira SE on a single node. Slowdown is caused mainly by Java Virtual Machine (JVM) overhead. 
We also wanted to identify the factors
that bound Kira SE's performance. For both Kira SE and the C implementation, we store data
locally in an ext4 file system.

\begin{figure}[h]
	\begin{center}
		\includegraphics[width=85mm]{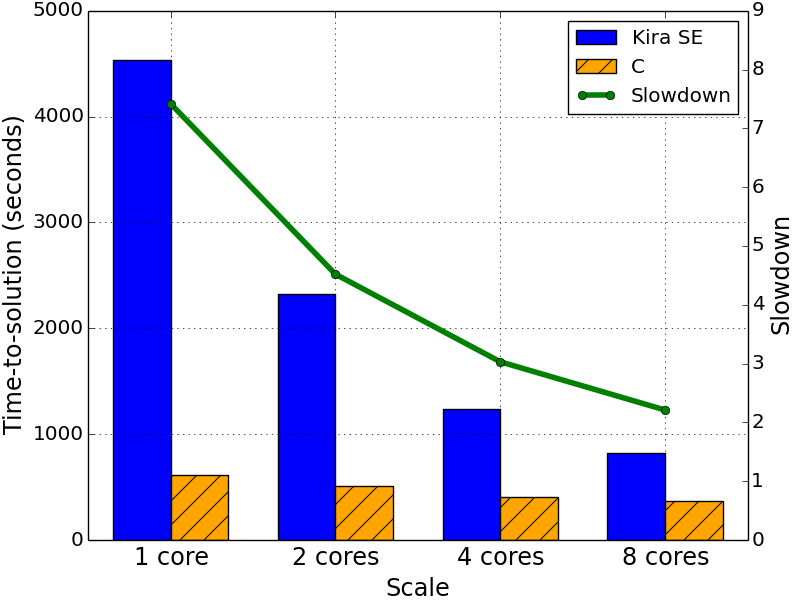}
		\caption{Single-Node Scale-Up Performance Comparison between Kira SE and the C Version (Lower is Better)
		\label{fig:scaleup}}
  	\end{center}
\end{figure}

For this experiment, we use a 12GB dataset from the SDSS DR2 survey. The dataset contains
2310 image files and each image is $\sim$6MB.
Figure~\ref{fig:scaleup} shows the relative
performance of Kira SE and the C version for various core counts on a single machine.

The C version is bound by local disk performance and shows limited improvement with
an increasing core count. While Kira SE is $7.4\times$ slower than the C implementation
on a single core, Kira SE is within $2.2\times$ the performance of the C version when using all eight cores on the node.
The scaling curves indicate that the performance of the approaches begins to converge
as core count increases. At convergence, both of the approaches will be saturating the local
disk bandwidth.

We also profile the C implementation with both warm and cold file system caches. When running with a
cold cache, the job completed in 371 seconds while the job completed in 83 seconds when
running with a warm cache. This indicates that 78\% of job execution time is consumed by
I/O (reading and writing data between local disk and memory). Since the C implementation of
SExtractor is disk bound, we believe that it is representative of a data intensive application. 

\subsection{Scale-Out Performance}
\label{sec:Performance-scaleout}

Next, we wanted to understand the strong scaling performance of both Kira SE and the C
implementation. Although Kira SE has 2.2--7.4$\times$ worse performance when running on
a single machine, we expect that Kira SE will achieve better performance at scale due
to disk locality. We expect that this will allow Kira SE to outperform the C implementation
on large clusters.

We use a 65GB dataset from the SDSS DR2 survey that comprises 11,150 image files.
Kira SE was configured to use HDFS as a storage system, while the C version used GlusterFS. 
Both HDFS and GlusterFS are configured with a replication factor of two.

\begin{figure}[h]
	\begin{center}
		\includegraphics[width=85mm]{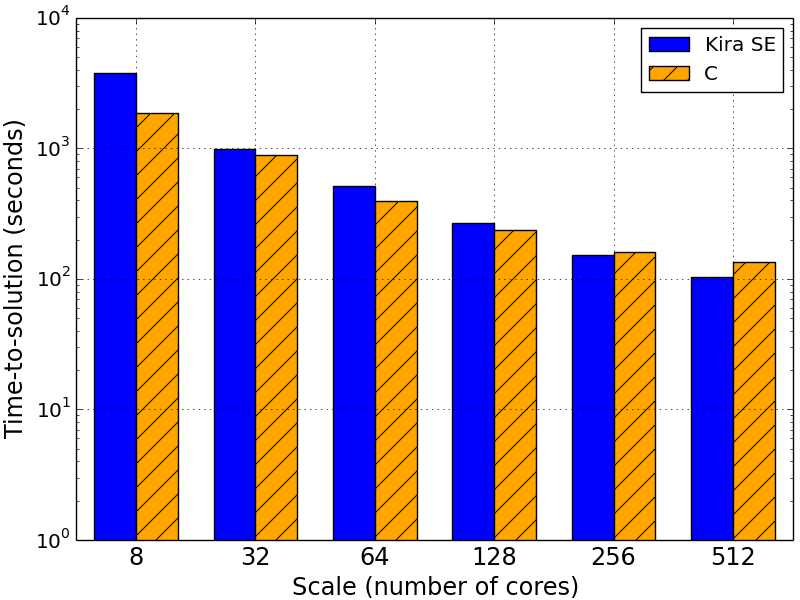}
		\caption{Scale-Out Performance Comparison between Kira SE and the C Version in Logarithmic Scale (Lower is Better)
		\label{fig:scaleout}}
  	\end{center}
\end{figure}

Figure~\ref{fig:scaleout} compares the performance of Kira SE and the C version across
multiple compute nodes in log scale. Kira SE is 2.7x slower as the C version
on eight cores. However, the gap between the two implementations decreases as we scale up. 
On 256 cores and 512 cores, Kira SE is respectively 5.6\% and 22.4\% faster than the C version.
Across all scales, Kira SE achieves near linear speedup.

\begin{figure}[h]
	\begin{center}
		\includegraphics[width=85mm]{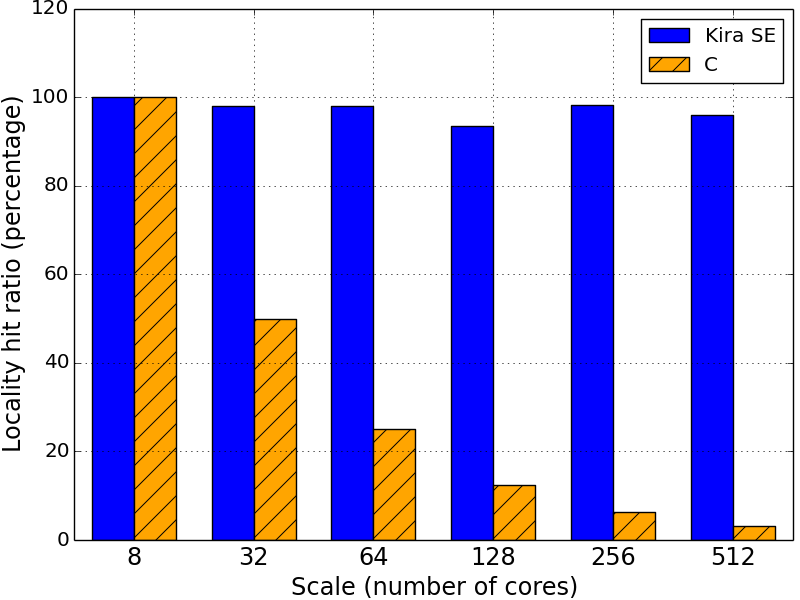}
		\caption{Locality Hit Ratio (Higher is Better)
		\label{fig:locality}}
  	\end{center}
\end{figure}

The fundamental driver of Kira SE's linear scalability is its consistent local disk
hit ratio, which is the ratio between the number of tasks that access the input file 
on local disk and total number of tasks. Spark and HDFS optimize for data locality during scheduling and achieve a
hit ratio above 98\%, as shown in Figure~\ref{fig:locality}. In contrast, the
C implementation's locality hit ratio decreases in half as the cluster size doubles.

In general, a shared file system can be configured in many ways to achieve better
availability, performance, and resilience. To understand the impact of the shared
file system configuration, we compare the performance of Kira SE (time-to-solution)
against four configurations of GlusterFS. The four configurations are
\emph{distributed}, \emph{replicated}, \emph{striped}, and \emph{striped replicated}. 
Table~\ref{tb:gluster-conf} explains the data layout of each configuration.
When possible, we set the replication and striping factors to two.
GlusterFS manages metadata in a distributed manner by spreading metadata across
all available nodes with a hash function. This allows the clients to deterministically
know the location of the metadata of a given file name in the cluster.

\begin{table}[h]
  \begin{center}
  \caption{GlusterFS Configuration Modes and Data Layout}
    \begin{small}
    \begin{tabular}{ | p{1.65cm} | p{6cm} |}
    \hline
    Configuration & Data Layout \\ \hline \hline
    distributed & files are distributed to all nodes without replication  \\ \hline
    replicated & files are distributed to all nodes with a number of replicas specified by the user \\ \hline  
    striped & files are partitioned into a pre-defined number of stripes then distributed to all nodes without replication \\ \hline
    striped replicated & files are partitioned into a pre-defined number of stripes and the stripes are distributed to all nodes with a number of replicas specified by the user \\ \hline
    \end{tabular}
    \end{small}   
  \label{tb:gluster-conf}     	
  \end{center}
\end{table}

\begin{figure}[h]
	\begin{center}
		\includegraphics[width=85mm]{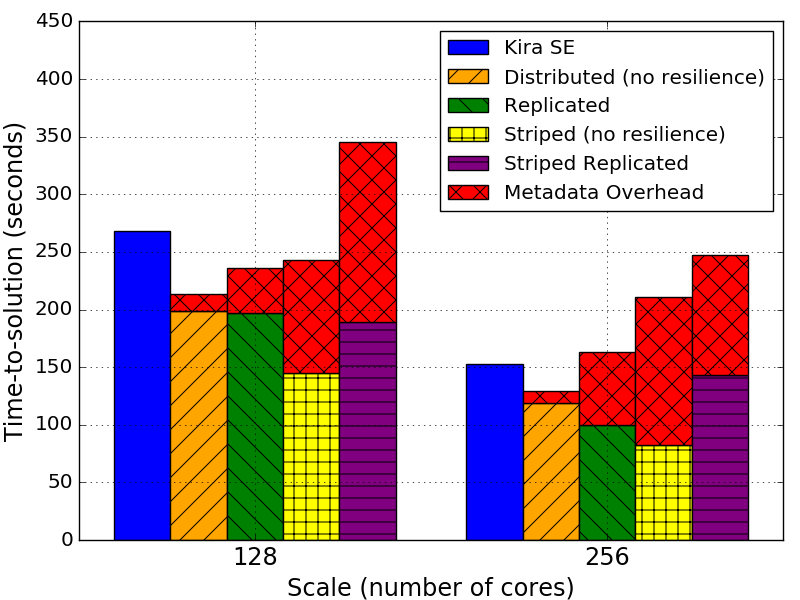}
		\caption{Kira SE Performance Compared to All GlusterFS Configurations (Lower is Better)
		\label{fig:allgluster}}
  	\end{center}
\end{figure}

We evaluate these configurations using the same dataset as the scale-out experiment. We
select 128 cores and 256 cores as the target scale since it is the transition point in
Figure~\ref{fig:scaleout} where Kira SE begins to outperform the C version. As stated
previously in~\S\ref{sec:HPC}, the C version performs a two-step process. The first step collects and
partitions all file paths. We refer to this step as metadata overhead. The processing
step occurs next, and is where each node processes its own partition.
Figure~\ref{fig:allgluster} compares the performance of Kira SE and the C version with
profiled metadata overhead.  

The C version running in the \emph{distributed} mode outperforms Kira SE at both scales. 
However, the \emph{distributed} mode is not practical in a cloud environment since
it has no replication or any other resilience mechanism. Replicating or striping
files introduces extra metadata overhead when we compare the \emph{replicated} mode
to the \emph{distributed} mode. The metadata overhead for each configuration increases
with large scales due to the cost of managing distributed metadata.

Another observation is that striping will further slow down metadata processing, whereas the
processing part takes less time than the \emph{distributed} mode for both scales due
to the doubled probability of accessing half file (with the striping factor of two) in local disk.
Since the input files are $\sim$6MB each, and are always processed by a single task, the
\emph{replicated} mode should be preferred to the \emph{striped replicated} mode.

On 256 cores, Kira SE outperforms all GlusterFS configurations except for the (impractical) \emph{distributed}
mode. Kira SE delivers comparable performance to this mode (18\% slower). In our
experiments with the 1TB dataset in Section~\ref{sec:1TB-EC2}, Kira SE outperforms
the \emph{distributed} mode.

\subsection{1TB Dataset Performance}
\label{sec:Performance-1TB}

We select a 1TB dataset from the SDSS DR7 survey, which is comprised of 176,938 image files. 
With this experiment, we seek to answer the following questions: 

\begin{itemize}
\item Can Kira scale to process a 1TB dataset?
\item What is the relative performance of Kira compared to the C version on EC2 resources?
\item How does Kira SE performance compare to that on a supercomputer?
\end{itemize}

\subsubsection{Cloud}
\label{sec:1TB-EC2}

\begin{figure}[t]
	\begin{center}
		\includegraphics[width=85mm]{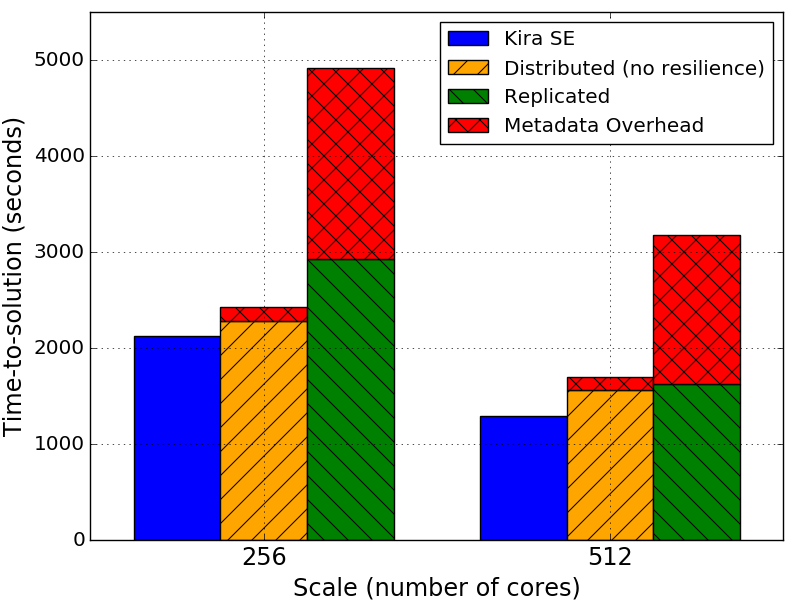}
		\caption{Kira SE Performance with 1TB Input Compared to the C Version Running on GlusterFS on EC2 (Lower is Better)
		\label{fig:1tb-ec2}}
  	\end{center}
\end{figure}

We configure GlusterFS in \emph{replicated} and \emph{distributed} modes and compare Kira
SE's performance against the C implementation. Kira SE runs $1.1\times$ and $1.3\times$ faster than the C version running on top of
GlusterFS configured in \emph{distributed} mode on 256 cores and 512 cores respectively. 
Using the more practical \emph{replicated} configuration of GlusterFS, Kira SE
is $2.3\times$ and $2.5\times$ faster. A detailed breakdown of the performance is shown in Figure~\ref{fig:1tb-ec2}.
The C version in \emph{distributed} mode is slower due to the node starvation that is introduced by our HPC solution.
The C version in \emph{replicated} mode slows down $2.2\times$ than that in \emph{distributed} mode because the directory metadata query
is dramatically slower ($13.4\times$), and the additional replica for each output file and associated metadata update introduces a slowdown
of $1.4\times$.


Compared to the experiment with the 65GB dataset in Section~\ref{sec:Performance-scaleout},
Kira SE processes $15.9\times$ more data in $12.5\times$ more time. Removing the Spark
startup time, we can see that Kira SE scales linearly in relation to the data size.

The overall throughput of Kira SE is 791MB/second, which is $2.4\times$ greater than necessary
to support the upcoming Large Synoptic Survey Telescope (LSST), as discussed in
Section~\ref{sec:Background-EngReq}. This high throughput enables real-time image processing.

\subsubsection{Supercomputer}

\begin{figure}[h]
	\begin{center}
		\includegraphics[width=85mm]{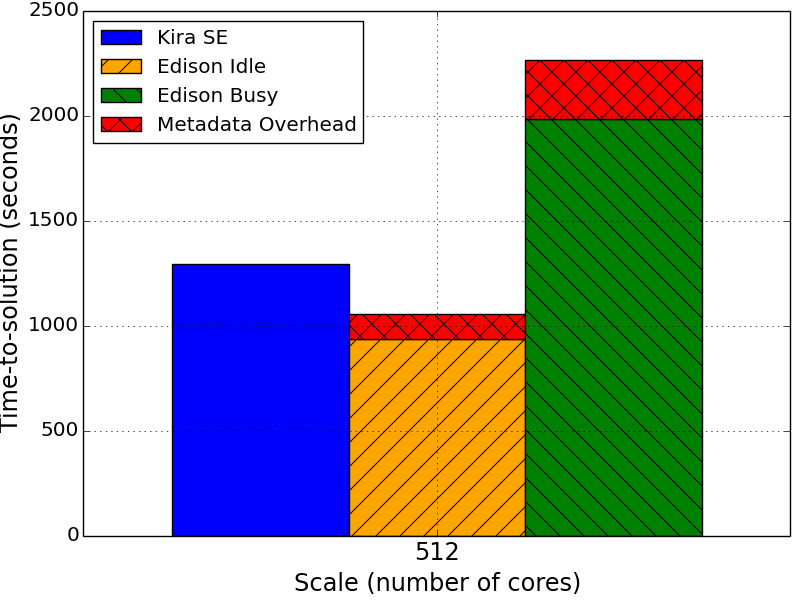}
		\caption{Kira SE Performance with 1TB Input Compared to the C Version Running on NERSC Edison Supercomputer (Lower is Better)
		\label{fig:1tb-edison}}
  	\end{center}
\end{figure}

Many astronomers have access to supercomputers and believe that
supercomputers outperform commodity clusters for data-intensive applications.
To examine this belief, we compare Kira SE on the Amazon cloud versus the performance of
the C version running on the NERSC Edison supercomputer, a Cray XC 30 System. We use
the Lustre file system which provides a peak throughput of 48GB/s. Each compute
node of Edison is equipped with a 24-core Ivy Bridge processor, with a 2.4GHz clock rate.
This is comparable to the CPU speed of the Amazon EC2 m2.4xlarge instance (eight vCPUs of
Intel Xeon E5-2665, each running at 2.4GHz). The experiments on Edison run on 21
nodes (a total of 504 cores) in the Cluster Compatibility Mode~(CCM)
while Kira SE uses 64 nodes (512 cores) on EC2. The Cray CCM provides a 
standard Linux cluster environment with services such as ssh, rsh, nscd, and ldap which 
are not supported in Cray's native mode.
Figure~\ref{fig:1tb-edison} shows the measurements.

During the experiments, we see that the C version performance varies significantly. These results
clearly fall into two classes. The first class has an average time-to-solution of 937.8 seconds
with a standard variation of 69.5 seconds. The second class has an average time-to-solution 
of 1983.2 seconds with a variation of 30.9 seconds. A further analysis shows that we are only 
using 0.4\% of the computing resources of the Edison machine. In the first class, the sustained
I/O bandwidth is 1.0 GB/s, which is 2.1\% of the I/O bandwidth available on the file system.
While in the second class, the sustained I/O bandwidth is down to 0.5GB/s.
Since the Edison cluster scheduler only schedules a single job per machine, computing resources 
are completely isolated. Thus, we can reason that, it is the I/O network resource or the file system 
that causes the performance variance. In Figure~\ref{fig:1tb-edison}, we use the term "idle mode" to refer to the 
situation where the I/O network is idle and the term "busy mode" to refer to the situation where the
I/O network is saturated. In general, Kira SE performs competitively as that of the C version
on Edison supercomputers.

\section{Related Work}
\label{sec:Related}

Many systems have tackled the problem of executing single process programs in parallel
across large compute clusters. This includes workflow systems such as HTCondor,
and ad hoc Hadoop and MPI based approaches.

In a workflow system, programmers can easily connect serial/parallel programs by specifying
dependencies between tasks and files. These systems do not require any modifications to
the original code base. Workflow systems provide a flexible data management and task execution
scheme that can be applied to a broad range of applications, but at the cost of programming
flexibility.

Researchers have used the Hadoop MapReduce~\cite{HADOOP} system to parallelize
tasks using a map-reduce data model. A variety of scientific applications have been parallelized
using Hadoop such as CloudBLAST~\cite{matsunaga08}. 
Although Hadoop exposes many convenient abstractions, it is difficult to express the application 
with the restrictive map-reduce API~\cite{dewitt08} and Hadoop's disk based model makes 
iterative/pipelined tasks expensive.

MPI has also been used to parallelize a diverse range of workloads. There
are MPI-based parallel implementations of astronomy image
mosaicing applications (Montage~\cite{jacob09}) and
sequence alignment and search toolkits (mpiBLAST~\cite{lin08}) applications. As an
execution system, MPI has two significant drawbacks. First, to implement a many-task
application on top of MPI, a user must develop a custom C wrapper for the application
and a custom message-passing approach for communicating between nodes. In practice, the
communication stages are critical for performance, which means that the
dataflow management scheme must be tailored to the application and hand tuned. Additionally,
MPI does not provide fault tolerance, which is problematic when running a long lived
application across many (possibly) unreliable nodes.

Traditionally, distributed workflow systems are run on top of a shared file system. 
Shared file systems (e.g., Lustre~\cite{donovan03}, and
GlusterFS~\cite{davies13}) are commonly used because they are compatible with the POSIX
standard and offer a shared namespace across all nodes. However, shared file systems
do not expose file locality to workflow systems, thus making suboptimal use of local
disks on the compute nodes when possible. Most tools in the Hadoop ecosystem use
HDFS~\cite{shvachko10}). HDFS  provides a shared namespace, but is not POSIX
compliant. Unlike traditional server-based shared file systems, HDFS uses
the disks on the compute nodes which enables data locality on filesystem access.

\section{Future Work}
\label{sec:Future}

The current Kira SE implementation is CPU bound. We plan to keep improving Kira SE's performance
on a single node until it is disk bound. This will enable Kira SE to make optimal use of
solid state drives~(SSDs) or in-memory file system that provide high I/O throughput.

Kira is currently available as an alpha release (\url{https://github.com/BIDS/Kira}), and we are planning an official 0.1 release
of the source extraction application in Fall 2015.
We plan to migrate to PySpark, which will enable better integration with other
Python-based astronomy tools. Additionally, we plan to provide several further enhancements to
the source extraction API.

Beyond the 0.1 release, we plan to integrate Kira with more astronomy image processing
programs, such as image reprojection and image co-addition. This will allow Kira to be
used as an end-to-end astronomy image analysis pipeline. We use this end-to-end pipeline
to continue evaluating the use of Spark as a conduit for many-task dataflow pipelines by
comparing against the equivalent C implementation. With this system, we will try to determine
which data intensive scientific applications execute most efficiently using ``big data''
software architectures on commodity clusters rather than using HPC software methods on supercomputers.
From this, we hope to obtain insights that can drive the development of novel computing
infrastructure for many-task scientific applications.

\section{Conclusion}
\label{sec:Conclusion}

In this paper, we investigated the idea of leveraging the modern big data platform for many-task
scientific applications. Specifically, we built Kira (\url{https://github.com/BIDS/Kira}), a flexible, scalable,
and performant astronomy image processing toolkit using Apache Spark running on Amazon EC2 Cloud. We also presented
the real world Kira Source Extractor application, and use this application to study the programming
flexibility, dataflow richness, scheduling capacity and performance of the surrounding ecosystem.

The Kira SE application demonstrates linear scalability with both increasing cluster and data
size. Due to its superior data locality, our Spark-based implementation delivers better performance than the equivalent C 
implementation running on GlusterFS. Specifically, Kira SE processes the 1TB SSDS DR7 dataset (176,938 tasks)
$2.5\times$ faster than C over GlusterFS when running on a cluster of 64 m2.4xlarge Amazon
EC2 instances. Kira SE also has comparable performance (between 18.5\% slower and 75.1\% faster)
to the C version running on the NERSC Edison supercomputer.  

We also demonstrated that Apache Spark can integrate with existing libraries.
This allows users to reuse existing source code to build new analysis pipelines.
A flexible interface, rich dataflow support, task scheduling capacity, locality optimization, and built-in support for fault tolerance make Spark a 
strong candidate to support many-task scientific applications. 
We experimented with Apache Spark as a popular example of a Big Data platform. We learned that
leveraging such a platform would enable scientists to benefit from the rapid pace of innovation 
and large range of systems and technologies that are being driven by wide-spread interest in Big Data analytics.

\section{Acknowledgments}

This research is supported in part by NSF CISE Expeditions Award CCF-1139158, DOE Award SN10040 DE-SC0012463, and DARPA XData Award FA8750-12-2-0331, and gifts from Amazon Web Services, Google, IBM, SAP, The Thomas and Stacey Siebel Foundation, Adatao, Adobe, Apple, Inc., Blue Goji, Bosch, C3Energy, Cisco, Cray, Cloudera, EMC2, Ericsson, Facebook, Guavus, HP, Huawei, Informatica, Intel, Microsoft, NetApp, Pivotal, Samsung, Schlumberger, Splunk, Virdata and VMware. Author Frank Austin Nothaft is supported by a National Science Foundation Graduate Research Fellowship.

This research is also supported in part by the Gordon and Betty Moore
Foundation and the Alfred P. Sloan Foundation together through the
Moore-Sloan Data Science Environment program.

This research used resources of the National Energy Research Scientific Computing Center, a DOE Office of Science User Facility supported by the Office of Science of the U.S. Department of Energy under Contract No. DE-AC02-05CH11231.
%
\bibliographystyle{abbrv}
\balance
\bibliography{Kira} 
%
%


\end{document}